\documentclass[aps,twocolumn,10pt,amssymb,nofootinbib,superscriptaddress,showkeys,showpacs,pra]{revtex4-1}
\usepackage{epsfig} 
\usepackage{amsmath} 
\usepackage{pifont}
\usepackage{graphicx}
\usepackage{color}
\usepackage{float}
\usepackage{varioref}
\usepackage{soul,xcolor}
\usepackage[export]{adjustbox}
\usepackage{multirow}
\usepackage[utf8]{inputenc}
\usepackage{tabularx,colortbl}
\usepackage{bbold}
\usepackage{braket}
\usepackage{titlesec}
\usepackage{placeins}
\usepackage{epstopdf}

\usepackage{latexsym}
\usepackage{dcolumn}
\makeatletter

\begin{document} 
	
	\title{Facets of quantum information under non-Markovian evolution}
	
	\author{Javid Naikoo}
	\email{naikoo.1@iitj.ac.in}
	\affiliation{Department of Physics, Indian Institute of Technology Jodhpur, Jodhpur, India}
	\author{Supriyo Dutta}
	\email{dosupriyo@gmail.com}
	\affiliation{Department of Theoretical Sciences, S. N. Bose National Centre for Basic Sciences, Salt Lake, Kolkata, India.}
	\author{Subhashish Banerjee}
	\email{subhashish@iitj.ac.in}
	\affiliation{Department of Physics, Indian Institute of Technology Jodhpur, Jodhpur, India}
	
	\begin{abstract}
		\noindent 
		\textbf{Abstract}
	 We consider two non-Markovian models:	Random Telegraph Noise (RTN) and non-Markovian dephasing (NMD). The memory in these models is studied from the perspective of quantum Fisher information flow. This is found to be consistent with the other well known witnesses of non-Markovianity. The two noise channels are characterized quantum information theoretically by studying their gate and channel fidelities. Further, the quantum coherence and its balance with mixedness is studied. This helps to put in perspective the role that the two noise channels can play in various facets of quantum information processing and quantum communication.	
	\keywords{}
	\end{abstract}
	
	\maketitle

	\section{Introduction}
		 The study of quantum systems interacting with their environment helps in characterizing the behavior of the dynamics of the system. This is useful in many application of quantum mechanics and has lead to the field of open quantum systems \cite{BP,SBbook}.
		 In many practical situations, the system-environment interaction brings in pronounced memory effects leading to the emergence of non-Markovian dynamics \cite{BreuerLainePiilo, rivas2010entanglement, RevModPhys.88.021002,RevModPhys.89.015001,li2018concepts,PhysRevA.62.042105,PhysRevE.67.056120}.  Recently, non-Markovianity has been a subject matter of various studies from quantum cryptography \cite{PhysRevA.83.042321,thapliyal2017quantum}, quantum biology \cite{lambert2013quantum,thorwart2009enhanced,huelga2013vibrations}, quantum metrology \cite{PhysRevLett.109.233601,Ban2015} and  quantum control \cite{PhysRevA.85.032321}. It has been shown with ample evidence that non-Markovian channels can be advantageous over Markovian ones. In \cite{bylicka2014non}, it was reported that the non-Markovianity can enhance the channel capacity in comparison to the Markovian case. Non-Markovian behavior is a multifaceted phenomenon which can not be attributed to a unique feature of the system-environment interaction. Consequently, several different measures were introduced in order to quantify the non-Markovian behavior, viz.,  trace distance~\cite{BreuerLainePiilo}, fidelity~\cite{vasile2011quantifying}, semigroup property \cite{wolf2008assessing} or divisibility \cite{rivas2010entanglement} of the dynamical map, quantum Fisher information (QFI) \cite{lu2010quantum}, quantum mutual information \cite{luo2012quantifying}. In general, these measures are inequivalent and different predictions by these measures have been reported in \cite{haikka2011comparing}.\par
		 The non-Markovian aspects become pertinent while dealing with quantum channels subjected to different types of environment. Another aspect of the system-environment interactions is the loss of the coherence and entanglement which is undesirable from the perspective of carrying out the tasks of quantum information. Therefore, this calls for the characterization of the quantum channels under the influence of different environments. Efforts have been made in this direction \cite{Omkar2013,PhysRevA.91.012324}. 
		 Quantum coherence can be thought of as a resource \cite{PhysRevLett.113.140401,PhysRevLett.116.120404,hu2018quantum} bringing out the utility of the quantum behavior in various tasks \cite{PhysRevLett.116.070402,PhysRevA.93.012110}. As the system evolves under ambient conditions, modeled by the noisy channel under consideration, it has a tendency of getting mixed \cite{banerjee2007dynamics}. A pertinent question then to ask is the trade-off between the mixedness and coherence \cite{PhysRevA.91.052115,dixit2018study}. The interplay between coherence and mixing in the context of non-Markovian evolution, has been studied \cite{bhattacharya2016evolution}. Gate fidelity \cite{nielsen2002simple}, which tell us  about the efficiency of the gate's performance and channel fidelity \cite{Omkar2013}, which is a measure of how well a gate preserves the distinguishability of states, and is thus connected to the Holevo bound of the channel, are two useful channel performance parameters. The performance of Lindbladian channels, such as the squeezed generalized amplitude damping (SGAD) channel \cite{PhysRevA.77.012318} and the Unruh channel \cite{banerjee2017characterization} have been studied using these parameters \cite{Omkar2013}. Here, these are used to characterize the two non-Markovian channels, RTN \cite{PhysRevA.70.010304,PradeepOSID,banerjee2017non} and NMD \cite{shrikant2018non}.\par
		   The non-Markovianity in these two channels can be witnessed by using QFI-flow. The contrast in the behavior of the dynamics in Markovian and non-Markovian regimes is highlighted in case of RTN.   The system considered is a general qubit state and the dynamics is given in terms of Kraus representation. This analysis would help to bring out the features of the system dynamics which plays a key role in the implementation of these two channels in quantum information and communication.\par
		 The paper is organized as follows: In Sec. (\ref{Facets}) we give a brief overview of quantum Fisher information as a witness of non-Markovianity and discuss various facets of quantum information. Section (\ref{DynamicsMaps}) is devoted to a brief discussion of the open system dynamics in terms of Kraus operators followed by a description of  RTN and NMD channels. In Sec. (\ref{QFIFnonMarkov}), we discuss the QFI-flow as a witness of non-Markovian behavior in the two channels. The results and their discussion is presented in Sec. (\ref{ResultsDiscussions}). We conclude in Sec. (\ref{Conclusion}).

	\section{Facets of quantum information}\label{Facets}
	 Here, we briefly describe various facets of quantum information used below to analyze the behavior of the dynamics of the RTN and NMD channels.\par
		\textit{Quantum Fisher Information {\rm (QFI)}}: Consider a $d$-dimensional quantum (qudit) state $\rho_\alpha$ depending on parameter $\alpha$. The QFI \cite{PhysRevA.91.052310,PhysRevA.88.043832} is a measure of the information with respect to the precision of estimating the
		inference parameter.
		   For the state parameter $\alpha$, the Fisher information is defined as
			\begin{equation}\label{fisher_info_in_gen}
	    	F_\alpha = \frac{\left[\vec{\zeta}(\alpha)\cdot \partial_\alpha \vec{\zeta}(\alpha) \right]^2}{1 - |\vec{\zeta}(\alpha)|^2} + \left| \vec{\zeta}_\alpha(\alpha) \right|^2.
	    	\end{equation}
	    	Here, $\vec{\zeta}(\alpha) = [\zeta_1(\alpha), \zeta_2(\alpha), \zeta_3(\alpha)]$ is  the Bloch vector for the general qubit state $\rho = \frac{1}{2} \mathbf{1} + \vec{\zeta}(\alpha) \cdot \vec{\sigma}$, with $\vec{\sigma}$ denoting the Pauli spin matrix triplet ($\sigma_x, \sigma_y, \sigma_z$).  One can define the QFI-flow as the time rate of change of the QFI as 
	    	\begin{equation}
	    	\mathcal{F}_\alpha = \frac{d F_\alpha}{dt}.
	    	\end{equation}
	    	
	    	In \cite{lu2010quantum}, it was proposed that a positive QFI-flow at time $t$ implies that the QFI flows back into the system from the environment, generating a non-Markovian dynamics. Therefore, we have
	    	\begin{equation}
	    	\mathcal{F}_\alpha = \begin{cases}
	    	< 0 &  {\rm Markovian~dynamics},\\\\
	    	> 0 &   {\rm non-Markovian~dynamics}.
	    	\end{cases} 
	    	\end{equation}
	    	The back flow of QFI is linked to the divisibility property of the underlying dynamical map. \par
	    	\textit{Quantum coherence:} Quantum coherence is a consequence  of quantum superposition and is necessary for existence of entanglement and other quantum correlations. The degree of quantum coherence in a state described by density matrix $\rho$ is given by its the off-diagonal elements. Specifically, the sum of the absolute values of the off-diagonal elements of $\rho$ serves as a measure of the coherence
	    	\begin{equation}\label{eq:Coh}
	    	C = \sum_{i\ne j} |\rho_{ij}|.
	    	\end{equation}
	    	The coherence parameter $C$  tends to zero with increase in mixing. \par
	    	\textit{Purity and Mixedness of quantum states}: For a normalized state $\rho$, the purity is a scalar quantity ${\rm Tr} \big[\rho^2\big]$ which is a measure of how much mixed a state is. Alternatively, one can define the mixedness parameter
	    	\begin{equation}
	    	\mathcal{M} = 2\bigg(1 - {\rm Tr} \big[\rho^2\big]\bigg),
	    	\end{equation}  
	    	such that $ \mathcal{M} = 0$ for pure state and $ \mathcal{M} = 2(1 - \frac{1}{d})$ for maximally mixed state. The interplay between coherence and mixedness was studied in  \cite{PhysRevA.91.052115}. For an arbitrary quantum state ($\rho$),  in $d$-dimensional Hilbert space,  the trade-off between coherence and mixing is quantified by the parameter $\beta$ given as:
	    	\begin{equation}\label{eq:beta}
	    	\beta = \frac{C^2}{(d-1)^2} + \mathcal{M} \le 1.
	    	\end{equation}
	    	\par
	    	\textit{Average gate fidelity}: One of the important tasks in quantum computation and quantum information is characterization of the quantum gates and quantum channels. In this direction, the average gate fidelity is a useful tool to quantify the quality of the quantum gates and is given by the compact expression  
	    	\begin{equation}
	    	G_{av} = \frac{1}{d(d+1)} \bigg(d + \sum\limits_{k}^{} {\rm Tr}\big[ E_k\big] \bigg).
	    	\end{equation}
	    	Here, $d$ is the dimension of the system and $E_k$ are the Kraus operators characterizing the quantum channel. It gives some idea of how well a quantum gate performs an operation it is supposed to implement.\par
	    	
	       \textit{Holevo information}:  Given any measurement described by the positive operator valued measure (POVM) $\{E_k\}$ performed on state $\rho = \sum_{i} p_i \rho_i$, we define the Holevo quantity as 
	       \begin{equation}
	       \chi_H = S(\rho) - \sum\limits_{i} p_i S(\rho_i).
	       \end{equation}
	       Holevo quantity represents the maximum amount of classical information that can be transmitted over a quantum channel.

	\section{Dynamics and maps}\label{DynamicsMaps}
	\begin{figure}
		\begin{center}
			\includegraphics[width=70mm]{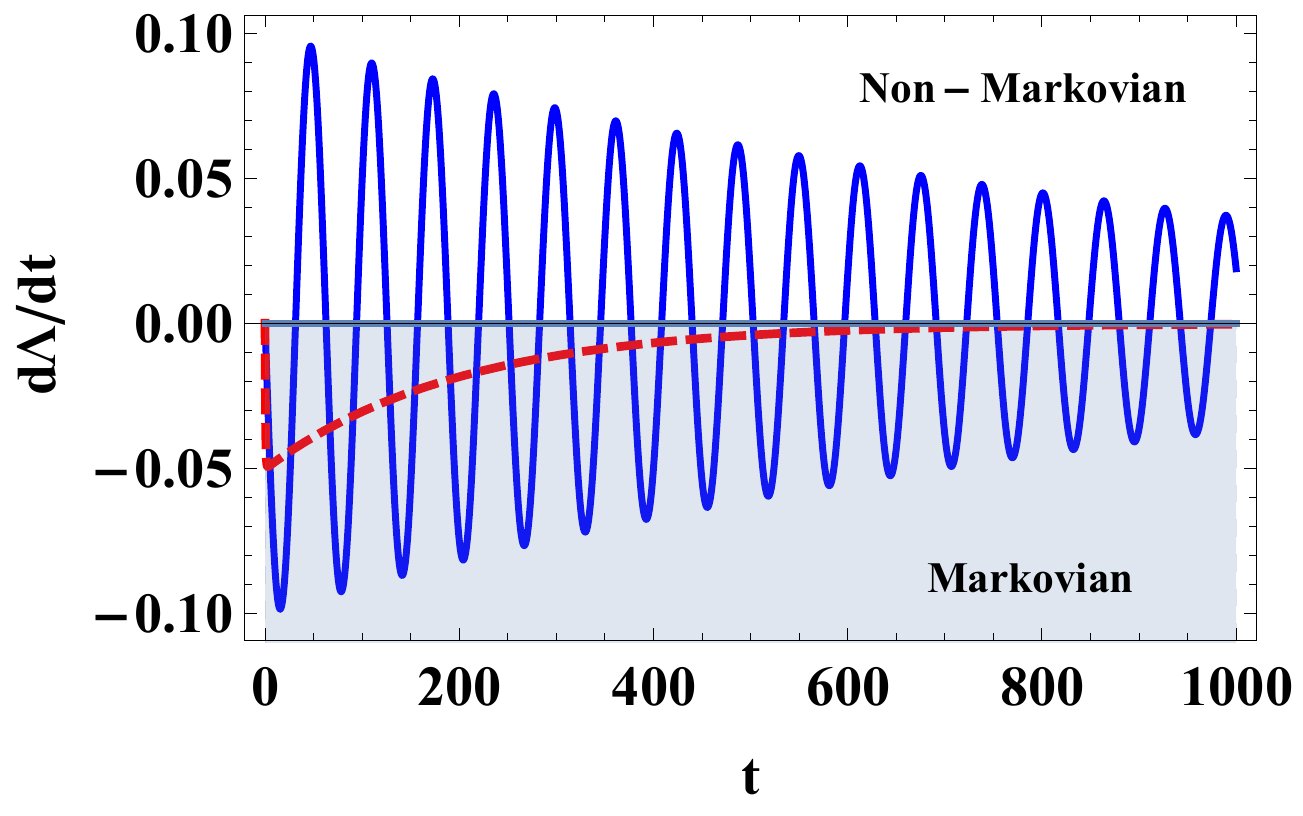}
			\includegraphics[width=70mm]{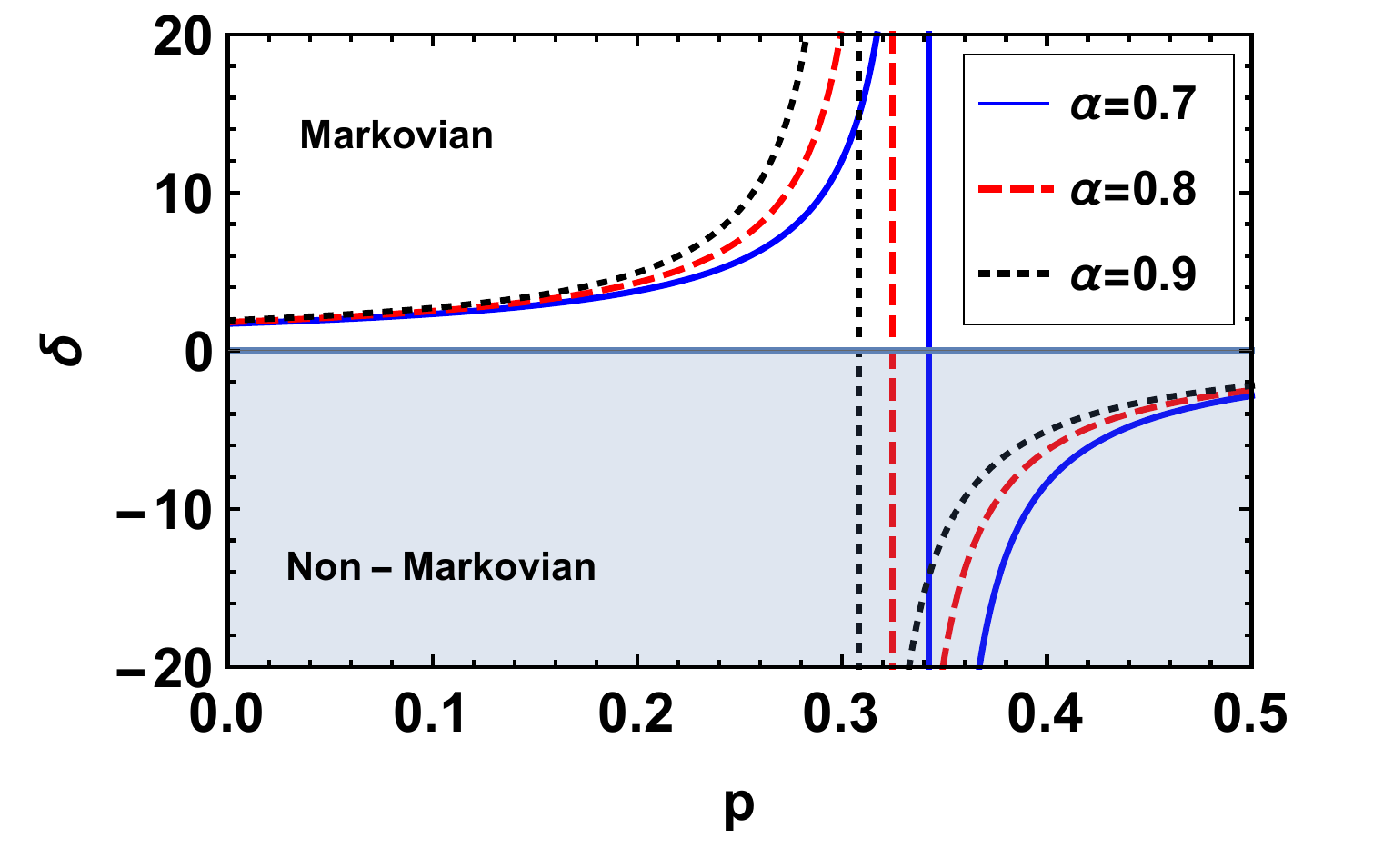}
		\end{center}
		\caption{Behavior of decoherence rate with respect to time $t$ for RTN (top) and with respect to $p$, a time like parameter, for NMD (bottom) channels, respectively. The non-Markovian dynamics is implied by $d\Lambda/dt > 0$ ($\delta < 0$). In case of RTN, the solid (blue) and dashed (red) curves correspond to non-Markovian ($a = 0.05, \gamma = 0.001$) and Markovian ($a = 0.05, \gamma = 1$) cases, respectively. Further, for RTN, the magnitude of the dashed (red) curve is increased ten times. The Markovian and non-Markovian regimes are separated by a singularity (vertical line) for the NMD channel.} 
		\label{fig:DecoRate}
	\end{figure}
     Now, we briefly review the Kraus representation of open system dynamics. This will be followed by the specific models like Random Telegraph Noise (RTN) and non-Markovian Dephasing (NMD).\par
		\textit{Kraus representation:} An open quantum system is not necessarily governed  by a unitary evolution, unlike a closed quantum system. A useful description for open quantum systems can be provided by Kraus representation \cite{kraus1983states}. The development of Kraus representation usually starts by assuming the system ($S$) and the environment ($E$) evolve together as a single system ($S+E$)  unitarily. One can then write the state for $S$, $E$ and $S+E$ as $\rho_S$, $\rho_E$ and $\rho_{SE}$, respectively.
		One can trace over the environment degrees of freedom and recover the evolution of the system alone
		\begin{equation}
		\rho_S(t) = {\rm Tr}_E\big[U_{SE}(t) \rho_{SE}(0) U_{SE}^\dagger (t) \big].
		\end{equation}
		Here, $U_{SE}(t)$ is the unitary governing the evolution of the combined system. If it is possible to express the above equation in the form 
		\begin{equation}
		\rho_S(t) = \sum\limits_{\mu} \mathcal{K}_\mu(t) \rho_S(0) \mathcal{K}_\mu^\dagger(t),
		\end{equation}
		subjected to the condition
		\begin{equation}
		\sum\limits_{\mu} \mathcal{K}_\mu(t) \mathcal{K}_\mu^\dagger(t) = \mathbf{I},
		\end{equation}
		we say that the evolution of $\rho_S(t)$ has the form of the Kraus representation. Such a representation is always possible if the system and environment do not share any correlation at $t=0$, i.e., if $\rho_{SE}(0) = \rho_{S}(0) \otimes \rho_{E}(0)$. However, as shown in \cite{Salgado}, the initial separability of states is not a necessary condition for Kraus representation.
         \begin{figure}
         	\begin{center} 
         		\includegraphics[width=70mm]{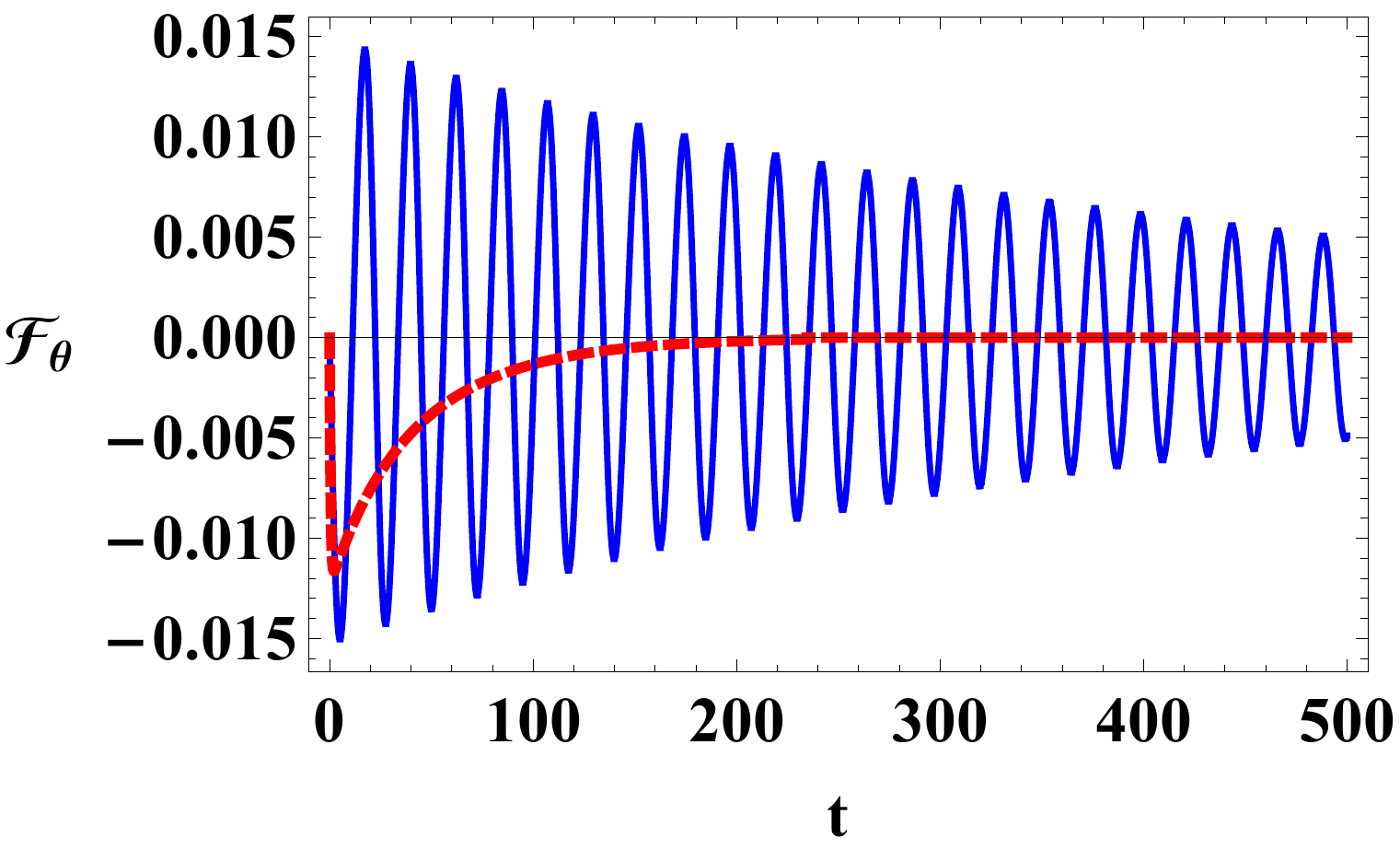}
         		\includegraphics[width=70mm]{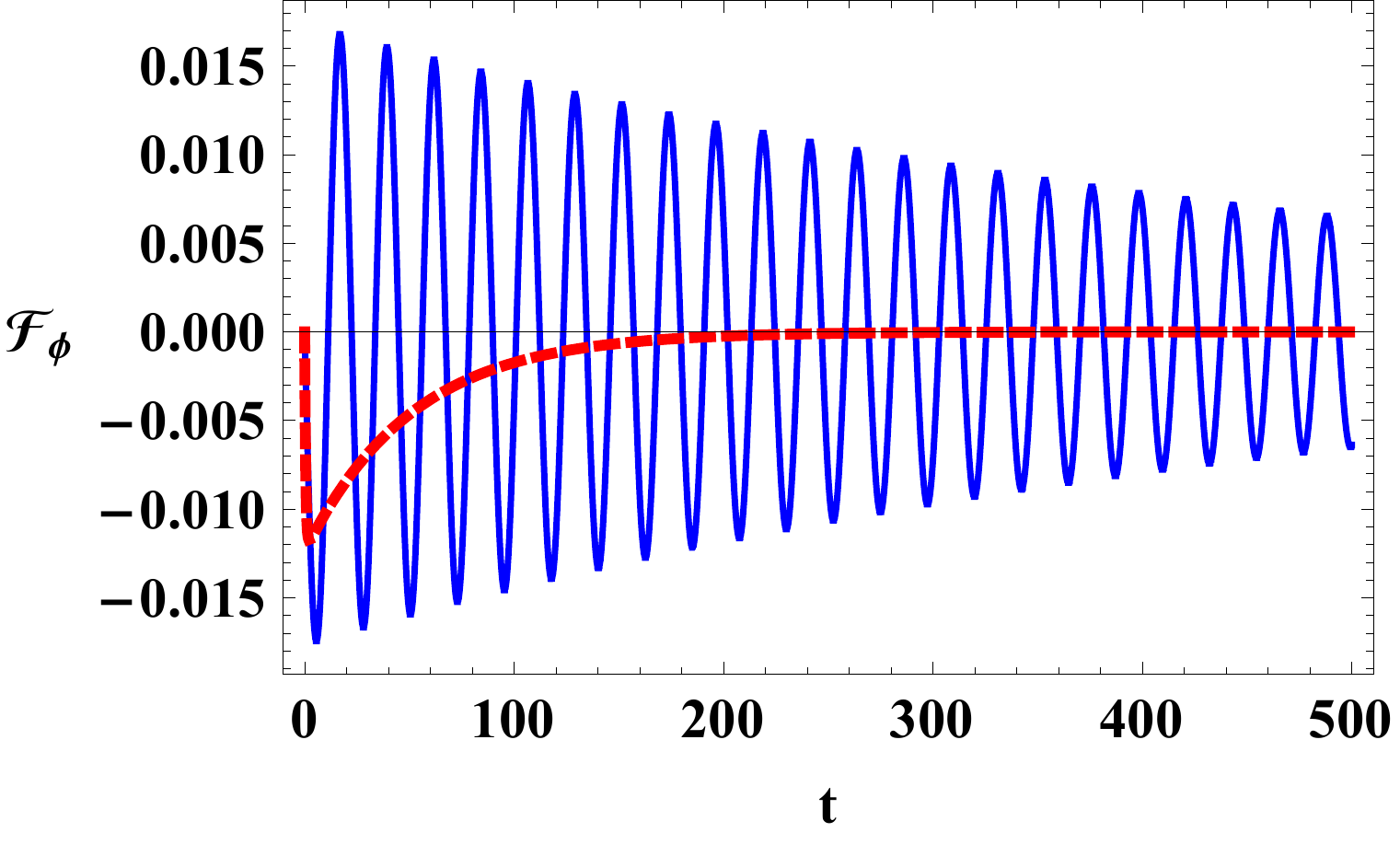}
         	\end{center}
         	\caption{RTN noise model: Fisher information flow corresponding to the parameters $\theta$ (top) and $\phi$ (bottom), as defined in Eqs. (\ref{fisher_rtn_theta}) and (\ref{fisher_rtn_phi}), respectively, are plotted with respect to time. Solid (blue) and dashed (red) curves correspond to the non-Markovian ($a = 0.07, \gamma = 0.001$) and Markovian ($a = 0.07, \gamma = 1$) cases, respectively. The state variable $\theta$ is chosen to be $\pi/4$. The magnitude of the dashed (red) curve is increased five times.} 
         	\label{fig:RTN_Fisher}
         \end{figure}

    \textit{Random Telegraph Noise:} The Random Telegraph Noise (RTN) is characterized by the autocorrelation function given as 
		\begin{equation}
		\langle \mathcal{X}(t) \mathcal{X}(s) \rangle = a^2 e^{-\gamma |t-s|},
		\end{equation}
		with $\mathcal{X}$ being the stochastic variable. The parameter $a$ is proportional to the system environment coupling strength and $\gamma$ controls the fluctuation rate of the RTN. The  map, $\mathcal{E}$, governing the time evolution under RTN  has the following Kraus representation
		\begin{equation}
		\mathcal{E}[\rho] = K_1 \rho K_1^\dagger + K_2 \rho K_2^\dagger,
		\end{equation}
		with
		\begin{equation}
		K_1(\nu) = \sqrt{\frac{1 + \Lambda(\nu)}{2}} \mathbf{I}, {\rm and}~ K_2(\nu) = \sqrt{\frac{1 - \Lambda(\nu)}{2}} \sigma_z.
		\end{equation}
		Here, $\Lambda(\nu) = e^{-\nu}[\cos(\nu \mu) + \frac{1}{\mu} \sin(\nu \mu)]$, with  $\mu = \sqrt{(\frac{2a}{\gamma})^2 - 1}$ and $\nu = \gamma t$. The dynamics is Markovian or non-Markovian depending on whether $(\frac{2a}{\gamma})^2 > 1$ or $(\frac{2a}{\gamma})^2 < 1$, respectively.

         Consider a general qubit state at time $t_0$ given as 
        \begin{equation}\label{qubit}
 			\rho = \begin{bmatrix} \cos^2 \frac{\theta}{2} & e^{-i\phi} \sin \frac{\theta}{2} \cos \frac{\theta}{2} \\ e^{i\phi} \sin \frac{\theta}{2} \cos \frac{\theta}{2} & \sin^2 \frac{\theta}{2}\end{bmatrix}.
 		\end{equation}
 		Under RTN noise, the state at some late time $t$ is given by
		\begin{equation}\label{eq:qubit_RTN}
		\begin{split} 
			\rho^\prime &=	\mathcal{E}_{t \leftarrow t_0}[\rho] = K_1 \rho K_1^\dagger + K_2 \rho K_2^\dagger \\
				        &= \frac{1 + \Lambda(t)}{2} \rho + \frac{1 - \Lambda(t)}{2} \sigma_3 \rho \sigma_3^\dagger \\
				        &= \begin{bmatrix} \cos^2 \frac{\theta}{2}             &   \frac{1}{2} e^{-i\phi} \sin \theta  \Lambda(t)\\
				              \frac{1}{2} e^{i\phi} \sin \theta  \Lambda(t)    & \sin^2 \frac{\theta}{2} \end{bmatrix}.
		\end{split}
		\end{equation}
	
	    Now, the dephasing master equation in its canonical from is given by
			\begin{equation}\label{eq:cannomaseqn}
			\dot{\rho} = \gamma (- \rho + \sigma_z \rho \sigma_z),
			\end{equation}
			 where $\gamma$ is the decoherence rate. The necessary and sufficient condition for a map to be CP-divisible is that the decoherence rate must be non-negative \cite{PhysRevA.89.042120}. Using Eqs. (\ref{eq:qubit_RTN})  and (\ref{eq:cannomaseqn}), the decoherence rate for the dephasing RTN map turns out to be  
		\begin{equation}
		\gamma = - \frac{1}{2 \Lambda} \frac{d \Lambda}{dt}.
		\end{equation}
		Since $\Lambda > 0$, the decoherence rate is negative when $\frac{d \Lambda}{dt}$ is positive. The negative decoherence rate is a signature of non-Markovian dynamics. As shown in Fig. (\ref{fig:DecoRate}), RTN shown negative decoherence rates for certain ranges of time $t$. This is consistent with the non-Markovian behavior studied using the QFI-flow, detailed below.
		\begin{figure}
			\begin{center}
				\includegraphics[width=70mm]{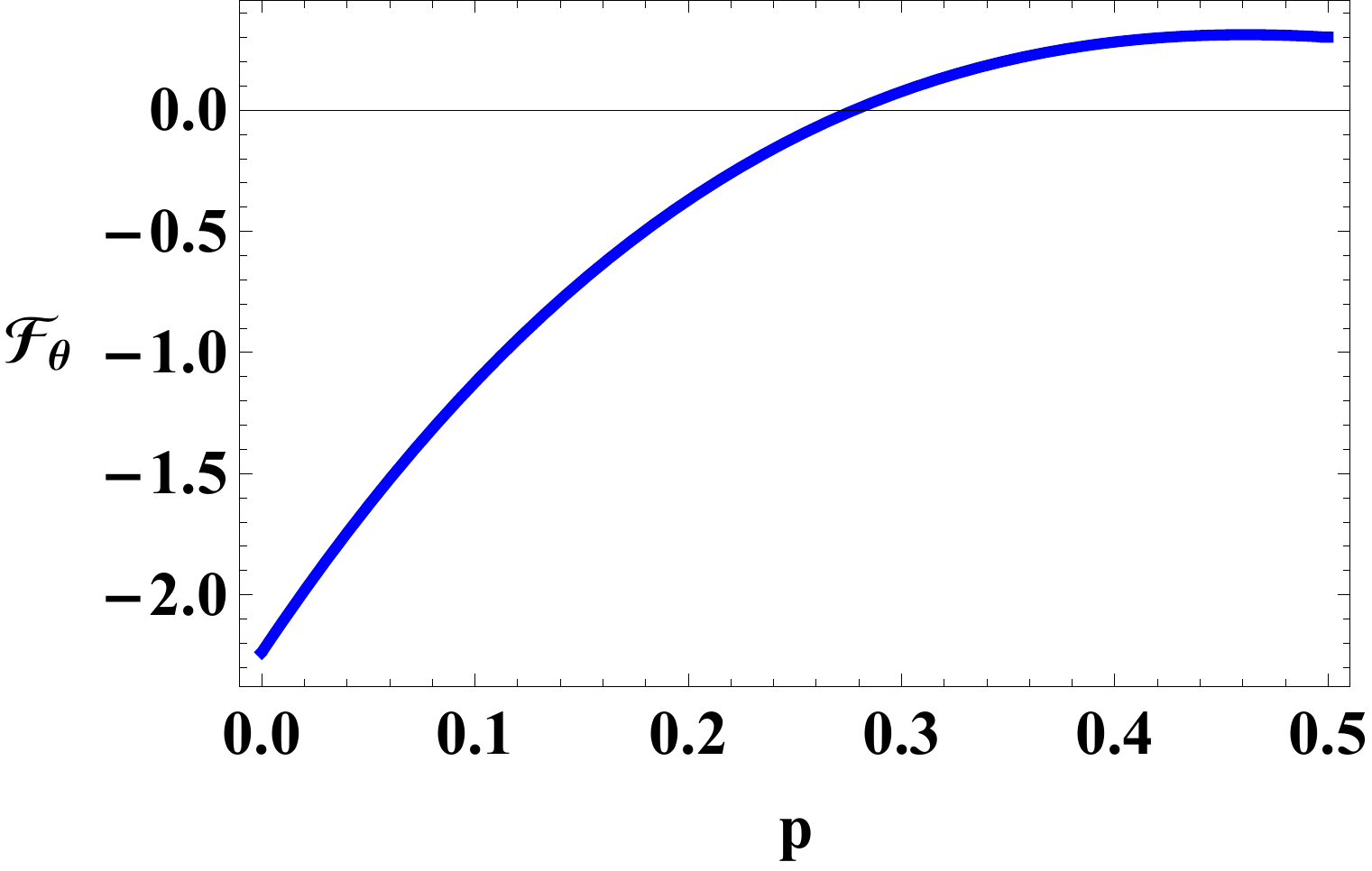}\\
				\includegraphics[width=70mm]{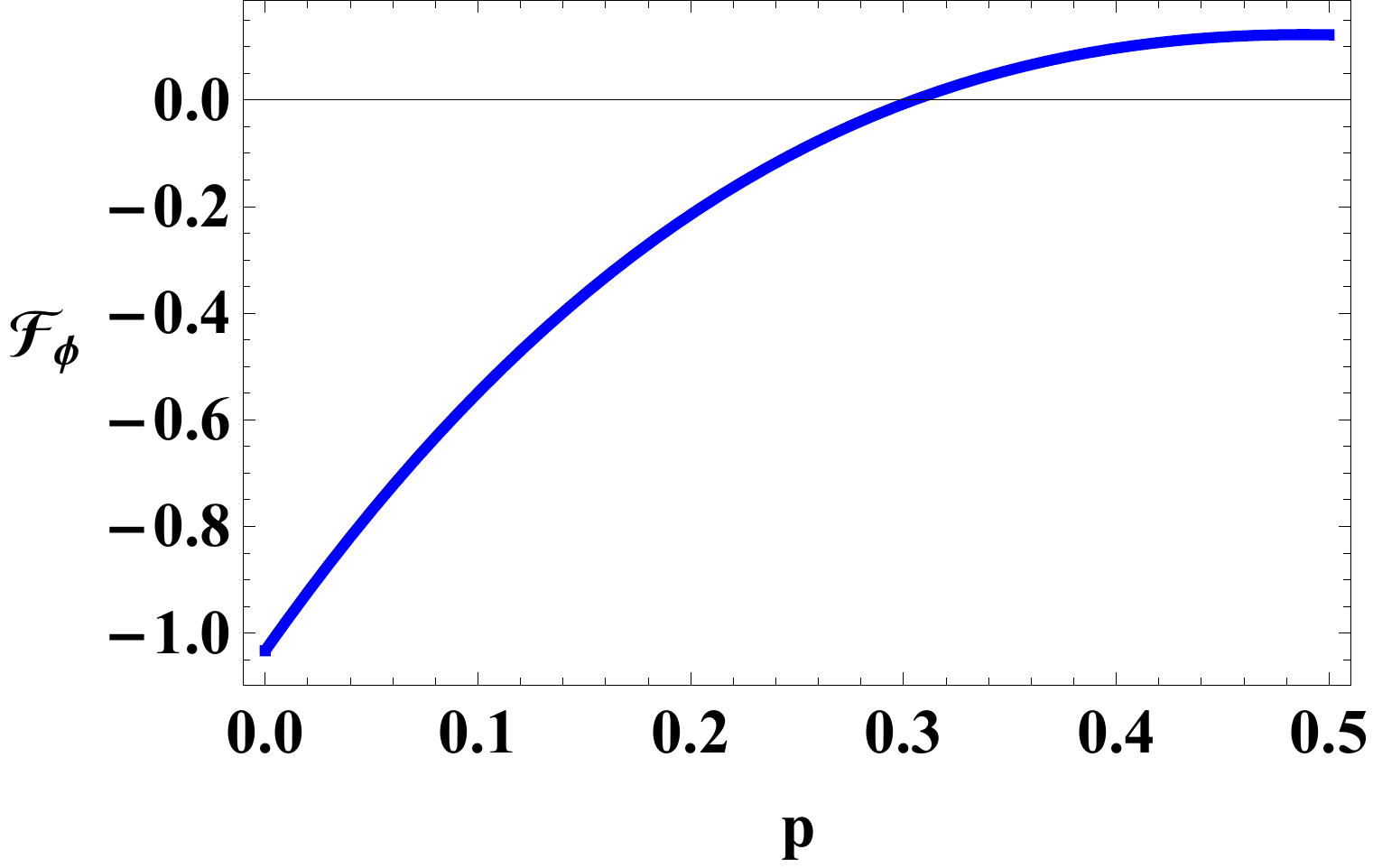}
			\end{center}
			\caption{Fisher information flow $\mathcal{F}_\alpha$ ($\alpha = \theta, \phi$), as defined in Eqs.(\ref{eq:NMDftheta}) and (\ref{eq:NMDfphi}), for non-Markovian dephasing dynamics. Non-Markovian dynamics is implied by $\mathcal{F}_\alpha > 0$. The parameters used are $\theta = \pi/4$ and $\alpha=0.7$.} 
			\label{fig:NMD_Fisher}
		\end{figure}
		
			\begin{figure}
			\begin{center}
				\includegraphics[width=70mm]{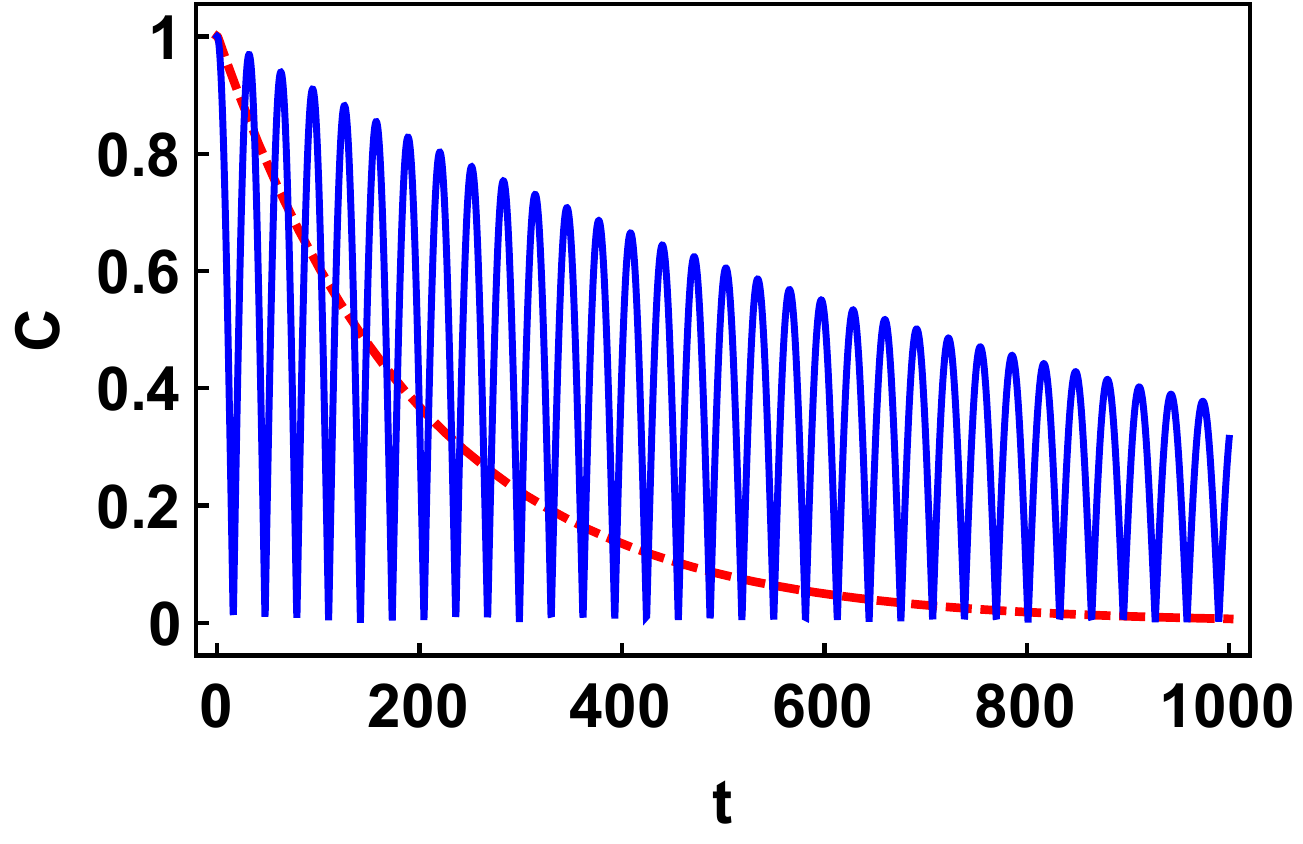}\\
				\includegraphics[width=70mm]{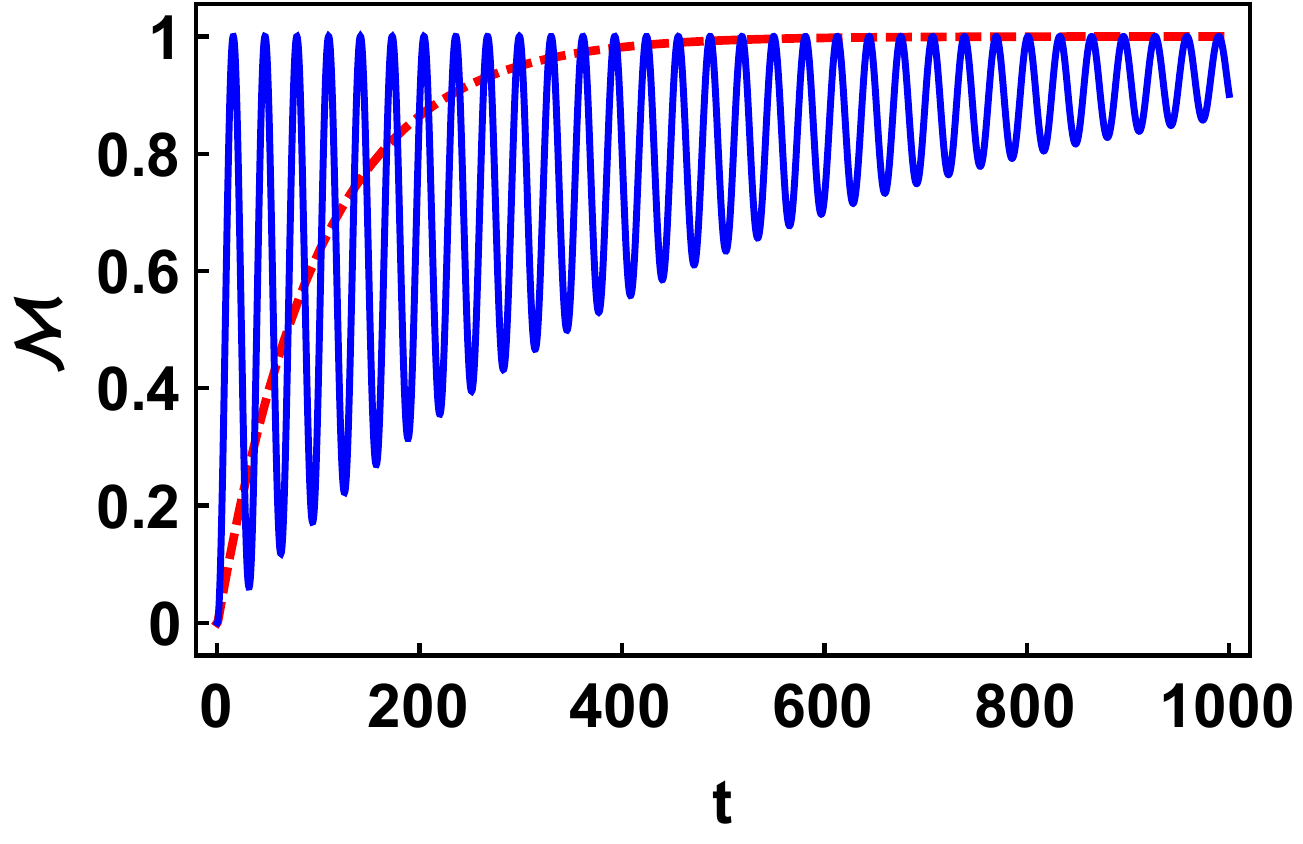}
			\end{center}
			\caption{Coherence parameter $C$ (top panel)  and mixedness parameter $\mathcal{M}$ (bottom panel) for a qubit under RTN evolution. The solid (blue) and dashed (red) curves correspond to non-Markovian and Markovian cases, respectively. Here, we have used $\theta = \pi/2$. The values of $a$ and $\gamma$ are the same as used in Fig. (\ref{fig:RTN_Fisher}).} 
			\label{fig:Coh_Mix_RTN}
		\end{figure}

	\begin{figure}
		\includegraphics[width=70mm]{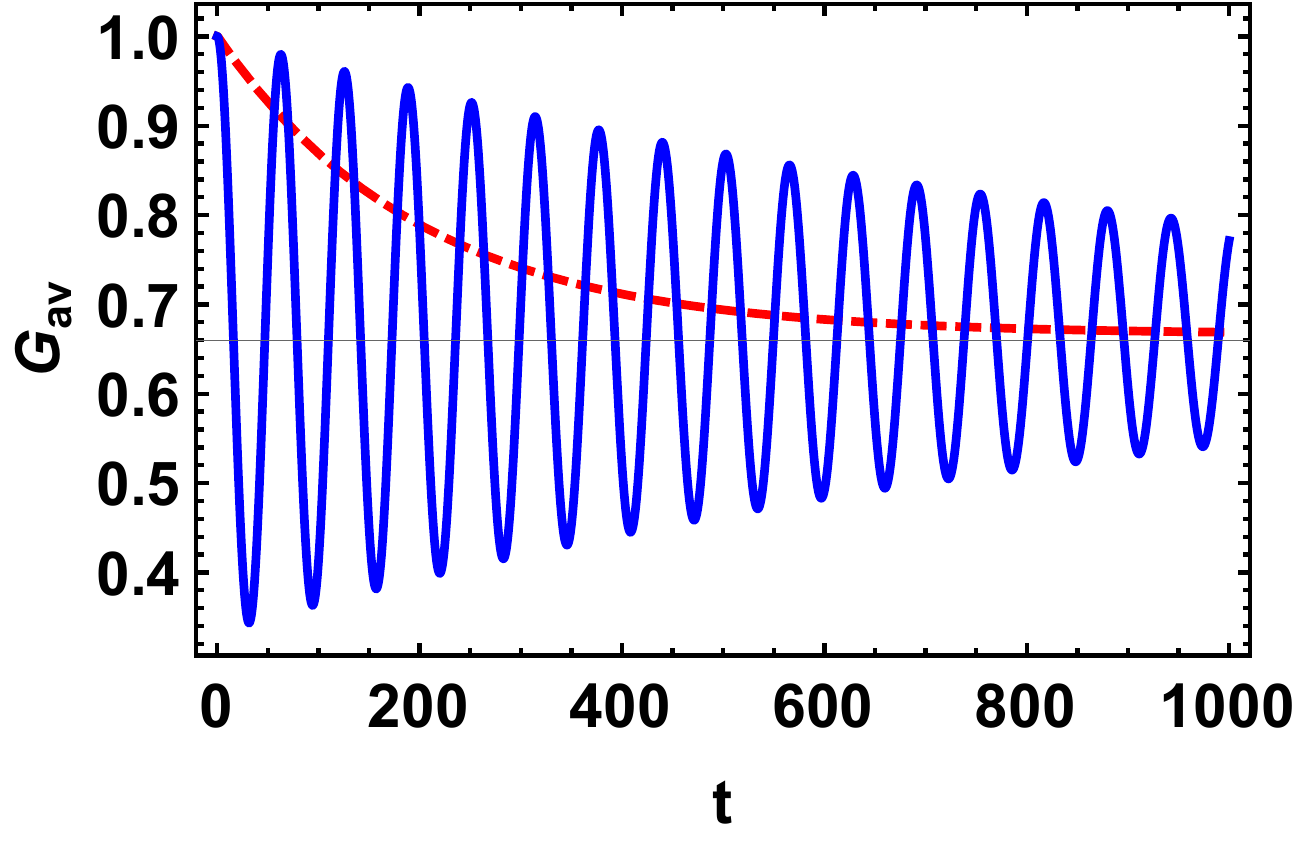}
		\includegraphics[width=70mm]{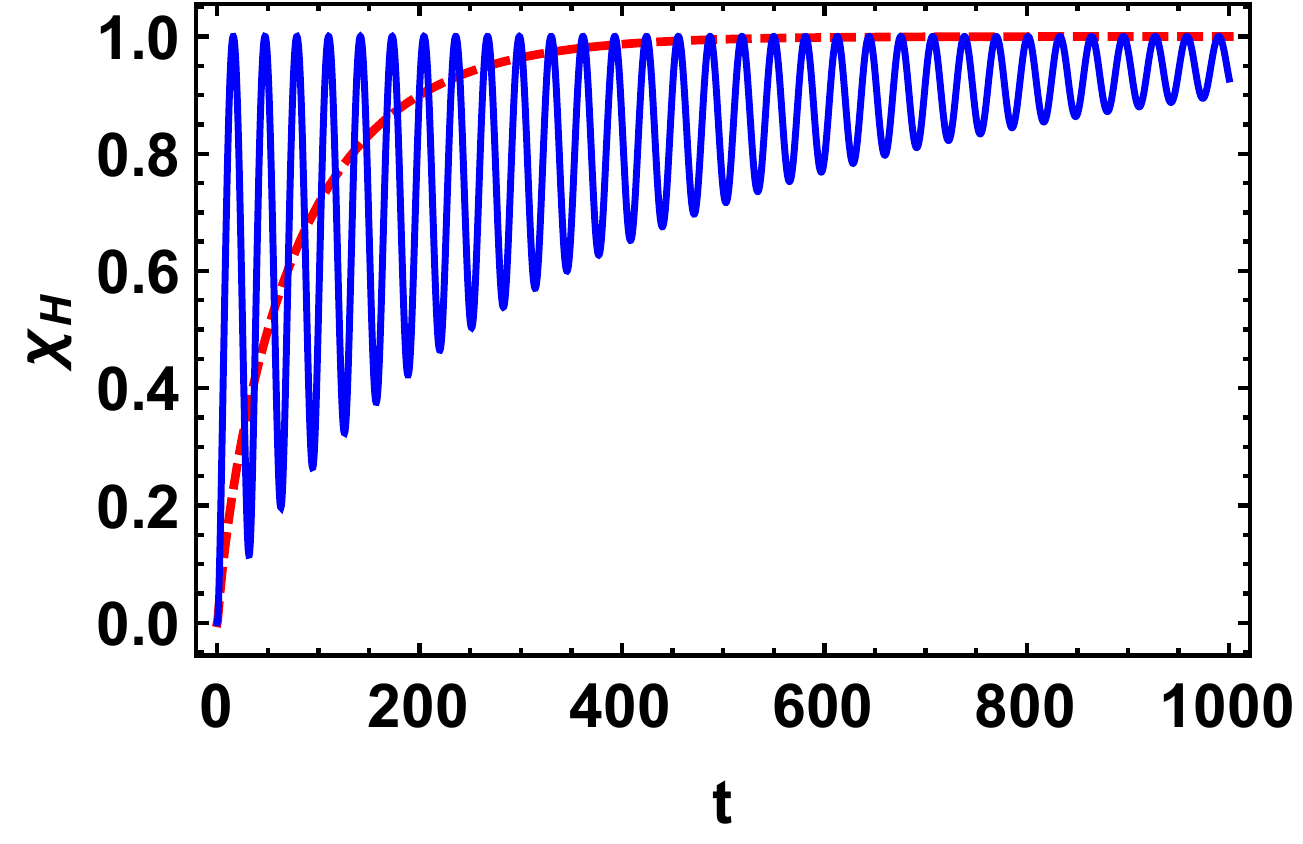}
		\caption{Average gate fedility $G_{av}$ parameter (top) and Holevo quantity $\chi_H$ (bottom) in RTN case. The solid (blue) and dashed (red) curves correspond to non-Markovian and Markovian cases, respectively. The parameters $a$ and $\gamma$ are as given in Fig. (\ref{fig:RTN_Fisher}). The state parameter $\theta = \pi/2$.} 
		\label{fig:Gfun_Holevo_RTN}
	\end{figure}	
	\textit{Non-Markovianian dephasing:}
	 The non-Markovina dephasing (NMD) is  governed by the following Kraus operators
	  \begin{equation}\label{krausNMD}
	  \begin{split} 
	  K_I & = \sqrt{[1-\alpha p](1-p)}~ \mathbf{I} = \sqrt{1 - \kappa}~ \mathbf{I},\\
	  K_z & = \sqrt{[1+ \alpha(1-p)]p}~ \sigma_z = \sqrt{\kappa}~ \sigma_z.
	  \end{split} 
	  \end{equation} 
    Here, $0 \le \alpha  \le 1$ and $p$ is a monotonically increasing function of time such that $0 \le p \le 1/2$. The above map reduces to  conventional dephasing in the limit $\alpha \rightarrow 0$. The action of the map, $\mathcal{E}^{ NMD}$, given by Kraus operators  in Eq. (\ref{krausNMD}), on a general qubit state in Eq. (\ref{qubit}) is 
	
     \begin{equation}
     \rho^\prime = \mathcal{E}^{NMD}_{t \leftarrow t_0} [\rho]
                 = \begin{bmatrix} \cos^2 \frac{\theta}{2}                       &     \frac{1}{2} e^{-i \phi } \sin (\theta ) \Omega \\ 
                 \frac{1}{2} e^{i \phi } \sin (\theta ) \Omega &      \sin^2 \frac{\theta}{2} 
                      \end{bmatrix}.\label{eq:qubit_NMD}
     \end{equation}
 Here, $\Omega = 1-2p+2p \alpha (p-1) = 1 - 2 \kappa$. Corresponding to the Kraus operators in Eq. (\ref{krausNMD}), the canonical master equation is 
 \begin{equation}\label{eq:cannonicalNMD}
 \dot{\rho}= \delta ( - \rho + \sigma_z \rho \sigma_z).
 \end{equation} 
 Here, $\dot{\rho} =\frac{d \rho}{dp}$ and the decoherence rate $\delta = \delta(p)$ as well as the state $\rho = \rho(p)$ are functions of the parameter $p$. Using Eqs. (\ref{eq:qubit_NMD}) and (\ref{eq:cannonicalNMD}), the decoherence rate turns out to be
 \begin{equation}
 \delta = \frac{\frac{1}{2}(r_+ + r_-) - p}{(p-r_-) (p-r_+)},
 \end{equation}
  with $r_{\pm}= (1 + \alpha \pm \sqrt{1 + \alpha^2})/2\alpha$. The regimes $p < r_-$ and $p > r_-$ correspond to Markovian and non-Markovian dynamics, respectively. The behavior of $\delta$ as a function of parameter $p$ is shown in Fig. (\ref{fig:DecoRate}). The singularity occurs at $p = r_-$, which, in turn, depends on the value of parameter $\alpha$.\par
 
\begin{table*}[htp]
	\centering
	\caption{\ttfamily Analytic expressions for various quantities studied in RTN and NMD models.}
{\ttfamily\begin{tabular}{|p{3cm}|p{7cm}|p{7.8cm}|} 
		\hline
		Facets $\downarrow$ Models$\rightarrow$\vspace{2mm}                  & RTN                                    &   NMD                                    \\
		\hline
		Coherence $(C)$ \vspace{2mm}                                         &  $\big|\Lambda \sin\theta \big| $      &     $ \big| (1-2\kappa) \sin (\theta ) \big|  $     \\
		\hline
		Mixedness $(\mathcal{M})$  \vspace{2mm}                              &  $(1- \Lambda^2)  \sin^2\theta$         &     $(1- (1-2\kappa)^2)  \sin^2\theta$             \\
		\hline
		  Coherence-Mixing~ \\balance $(\beta)$ \vspace{2mm}                 & $\sin^2\theta$                           &   $  \sin^2 \theta  $ \\
		\hline
		Average gate fedility \vspace{2mm}                                   & $(1 + |1+ \Lambda|)/3$                  &    $(1 + |1+ (1-2\kappa)|)/3$ \\
		\hline
		Holevo information  \vspace{2mm}                                     & $-\lambda_{+}{\rm Log}_2\lambda_{+}-\lambda_{-}{\rm Log}_2\lambda_{-}$            &                   $\frac{4 p (1-3 \alpha  (p-1)) \tanh ^{-1}\left(4 \alpha  (p-1) p-\frac{4 p}{3}+1\right)-3 \ln \left(2 \alpha  (p-1) p-\frac{2 p}{3}+1\right)}{\ln (8)}$ \\
		\hline
	\end{tabular}\label{table}
	Here $\lambda_{\pm} = \frac{1}{4} \left(2 \pm \sqrt{2} \sqrt{1 + \Lambda ^2 +  (1 -\Lambda ^2+) \cos2\theta} \right)$, $\Lambda =  e^{-\nu}[\cos(\nu \mu) + \frac{1}{\mu} \sin(\nu \mu)]$ and $\kappa = p[1 + \alpha  (1-p)]$. Also, $\theta$ is the state parameter, Eq. (\ref{qubit}).}
\end{table*}

\section{Quantum Fisher infomation flow and non-Markovianity}\label{QFIFnonMarkov}
In this section, we discuss the interplay between QFI-flow and non-Markovianity in the context of RTN and NMD channels by using the dynamics sketched in the previous section.\par
\textit{For RTN channel}:	We will use the time evolved state given in Eq. (\ref{eq:qubit_RTN}) and compute the QFI and QFI-flow. The Bloch vector corresponding to $\rho^\prime$ in Eq. (\ref{eq:qubit_RTN}) turns out to be 
\begin{equation}
\vec{\zeta} (\theta, \phi) =  \bigg( \frac{1}{2} \Lambda(\nu) \sin\theta \cos\phi,  \frac{1}{2} \Lambda(\nu) \sin\theta\sin\phi,  \frac{1}{2} \cos\theta\bigg).
\end{equation}		
Therefore,
\begin{align*}
\partial_\theta \vec{\zeta} (\theta, \phi) &= \left(\frac{1}{2} \Lambda(\nu) \cos\theta \cos\phi, \frac{1}{2} \Lambda(\nu) \cos\theta\sin\phi, - \frac{1}{2} \sin\theta \right), \\ 
\partial_\phi \vec{\zeta} (\theta, \phi) &= \left(- \frac{1}{2} \Lambda(\nu) \sin\theta \sin\phi, \frac{1}{2} \Lambda(\nu) \sin\theta \cos\phi, 0 \right).
\end{align*}
Also,

\begin{align}
& \vec{\zeta} (\theta, \phi) \cdot \partial_\phi\vec{\zeta} (\theta, \phi) = 0,\nonumber\\
& \vec{\zeta} (\theta, \phi) \cdot \partial_\theta\vec{\zeta} (\theta, \phi) = \frac{1}{4}((\Lambda(\nu))^2 - 1) \sin\theta \cos\theta,\nonumber \\		
& |\vec{\zeta} (\theta, \phi)|^2 =  \frac{1}{4}(\Lambda(\nu))^2 \sin^2(\theta) + \frac{1}{4}\cos^2\theta.
\end{align}

With above setting, the QFI corresponding to the parameters $\theta$ and $\phi$ becomes
\begin{equation}
\begin{split} 
F_\theta &= 1 + \frac{3 (-4 + \Lambda^2)}{2 (7 - \Lambda^2 + (-1 + \Lambda^2) \cos2 \theta)},\\
F_\phi   &= \frac{1}{4} \Lambda^2 \sin^2\theta.
\end{split} 
\end{equation} 

The corresponding  QFI-flows are given by the following expressions 

\begin{widetext}
\begin{align}
\mathcal{F}_{\theta}  &= \frac{dF_\theta}{dt} = -\frac{18 \gamma  \mu ^2 \left(\mu ^2+1\right) \cos ^2(\theta ) e^{2 \gamma  t} \sin (\gamma  \mu  t) \mu \Lambda(t)}{\big[\mu ^2 (\cos (2 \theta )-7) e^{2 \gamma  t}+2 \sin ^2(\theta ) \mu^2 \Lambda^2(t)\big]^2},\label{fisher_rtn_theta}\\
\mathcal{F}_{\phi}   &= \frac{dF_\phi}{dt} = \frac{1}{2}\sin^2\theta \frac{d\Lambda}{dt} = \frac{\gamma  \left(\mu ^2+1\right) \sin ^2(\theta ) e^{-2 \gamma  t} \sin ^2(\gamma  \mu  t) (\mu  \cot (\gamma  \mu  t)+1)}{2 \mu ^2}. \label{fisher_rtn_phi}
\end{align}
\end{widetext}

These quantities are depicted in the Fig. (\ref{fig:RTN_Fisher}) both for the Markovian as well as the non-Markvoian cases. \par

\textit{For NMD channel}: The analytic expressions for the QFI-flow in this case are given as
\begin{widetext}
\begin{align}
\mathcal{F}_\theta &= \frac{9 \cos ^2(\theta ) (2 \alpha  (p-1) p-2 p+1) (\alpha  (2 p-1)-1)}{(2 (p-1) p \cos (2 \theta ) (\alpha  (p-1)-1) (\alpha  p-1)-2 (p-1) p (\alpha  (p-1)-1) (\alpha  p-1)+3)^2}, \label{eq:NMDftheta}\\ \nonumber \\
\mathcal{F}_\phi &= \sin ^2(\theta ) (2 \alpha  (p-1) p-2 p+1) (\alpha  (2 p-1)-1).\label{eq:NMDfphi}
\end{align}
\end{widetext}

These quantities are plotted in Fig. (\ref{fig:NMD_Fisher}). The various facets studied in RTN and NMD models are listed in Table (\ref{table}) with their compact analytic expressions.

	\begin{figure}
	\includegraphics[width=70mm]{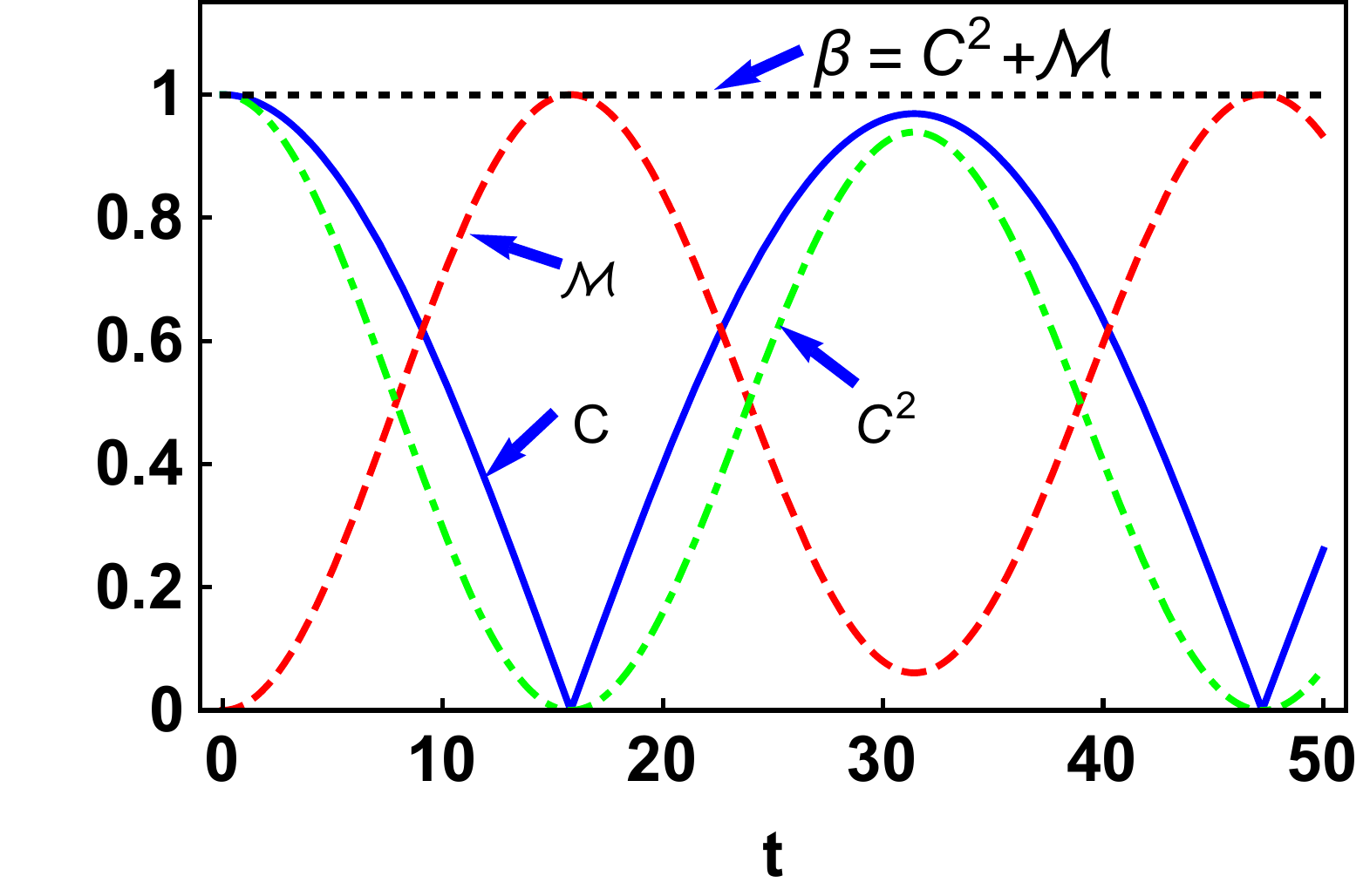}
	\includegraphics[width=70mm]{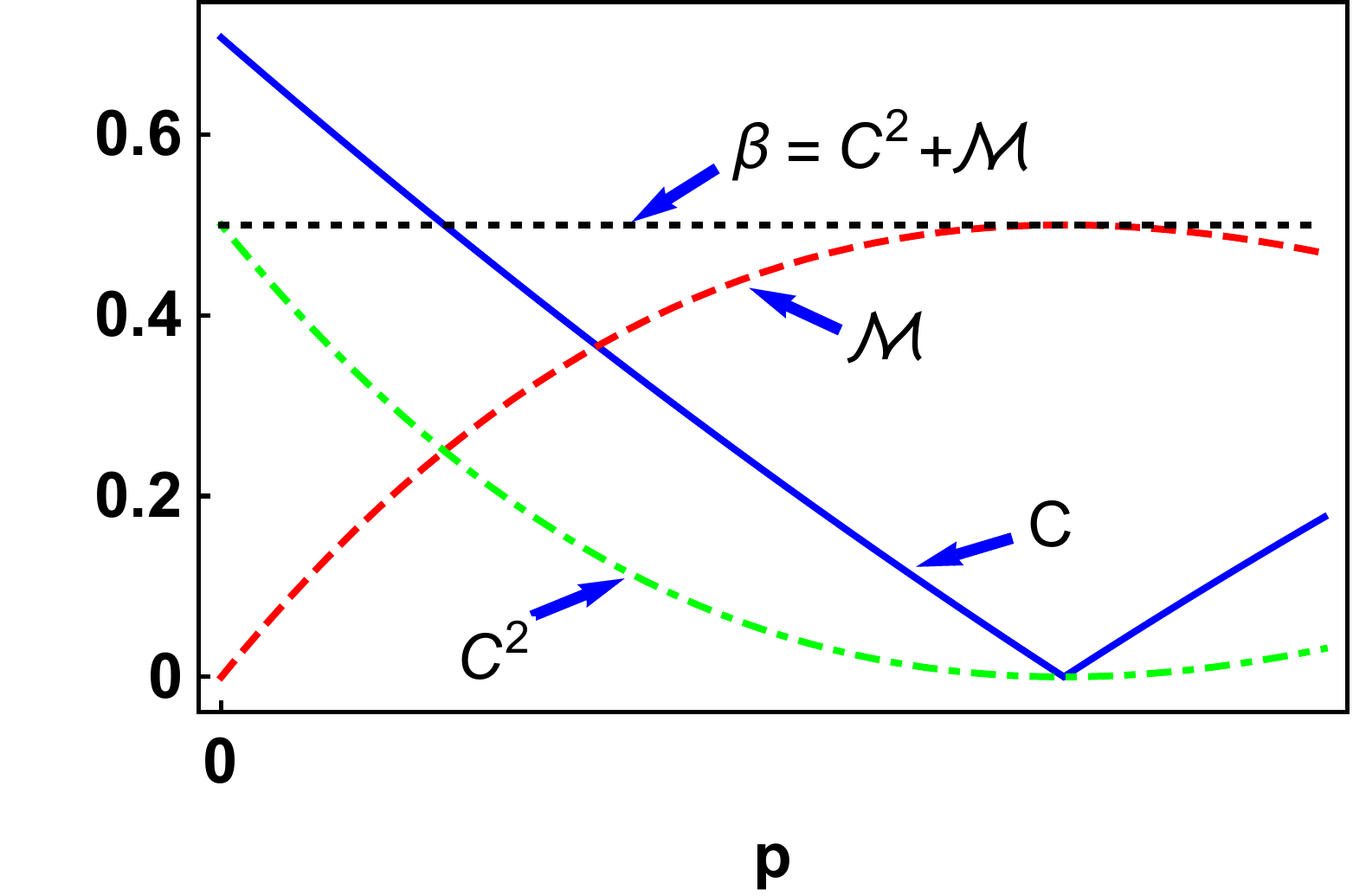}
	\caption{Interplay between coherence ($C$) and mixedness ($\mathcal{M}$) quantified by the parameter $\beta = C^2 + \mathcal{M}$, Eq. (\ref{eq:beta}). Top and bottom panel corresponds to RTN and NMD cases, respectively. For RTN, $a=0.5$, $\gamma = 0.001$ (non-Markovian case, similar trade-off is observed in Markovian case, not displayed here) and $\theta = \pi/2$, while for NMD the channel, $\alpha =0.5$ and $\theta =\pi/4$.} 
	\label{fig:Coh_Mix_Interplay}
\end{figure}

\begin{figure*}
	\begin{center}
		\includegraphics[width=43mm]{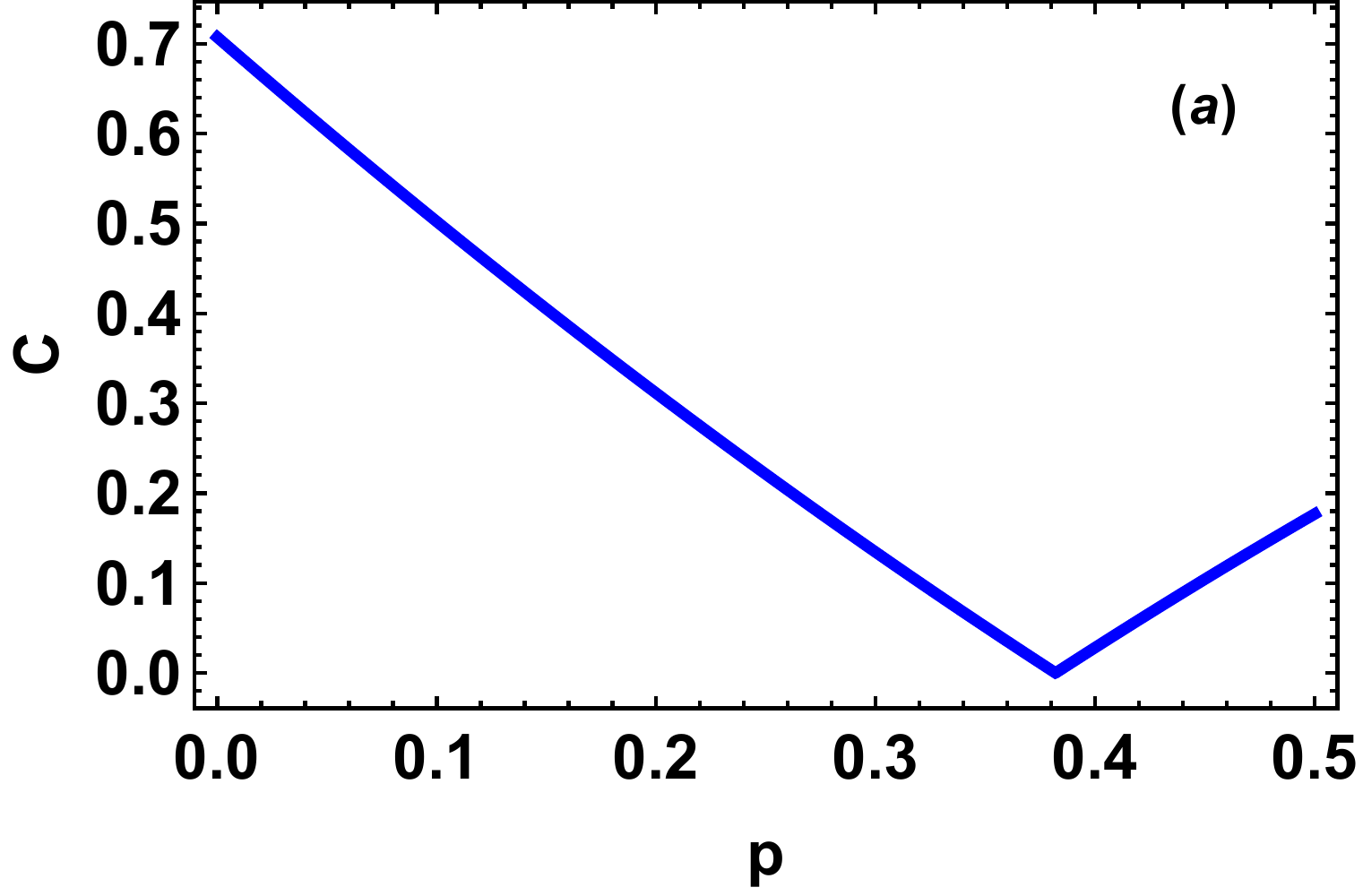}
		\includegraphics[width=43mm]{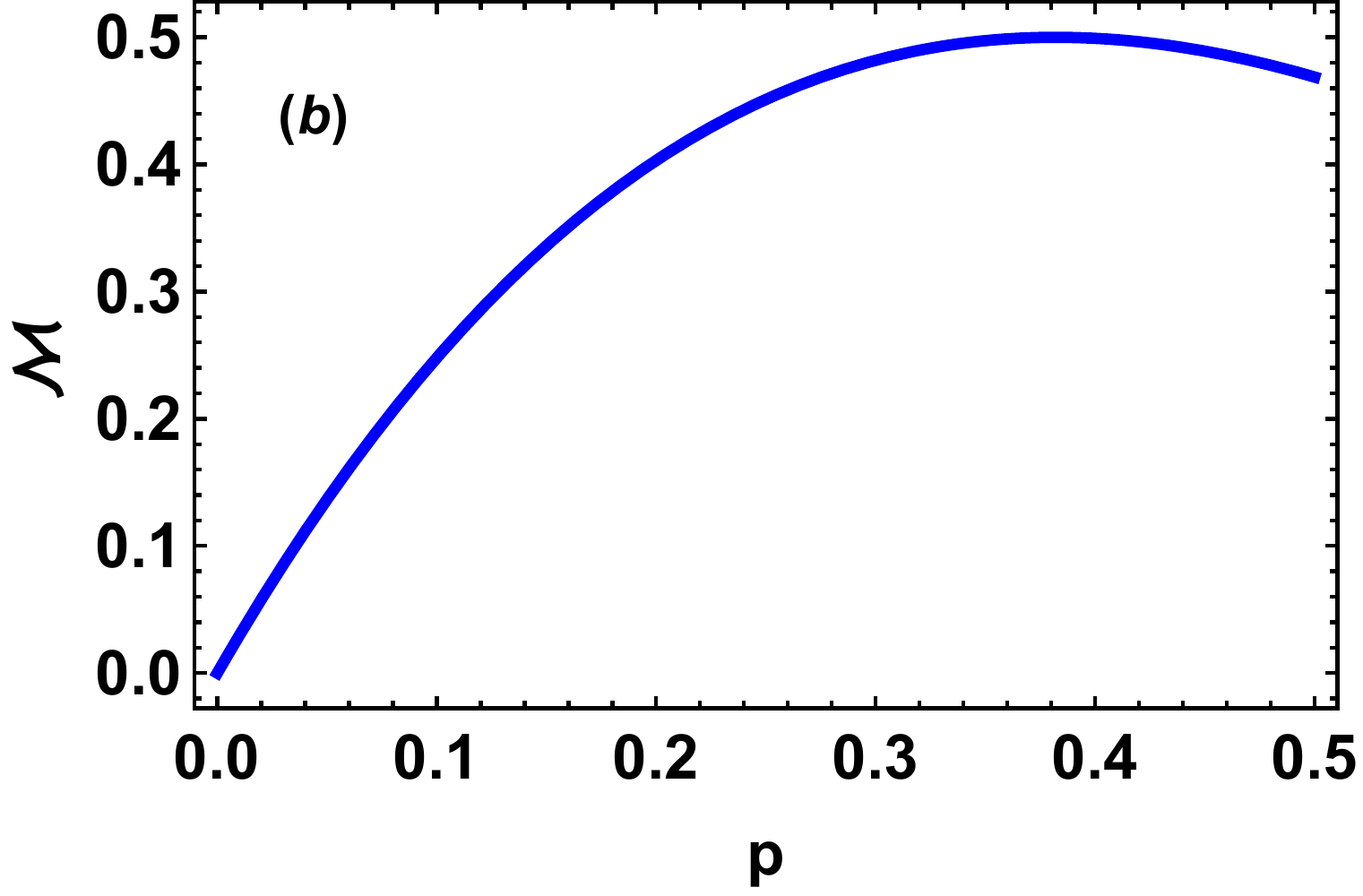}
		\includegraphics[width=43mm]{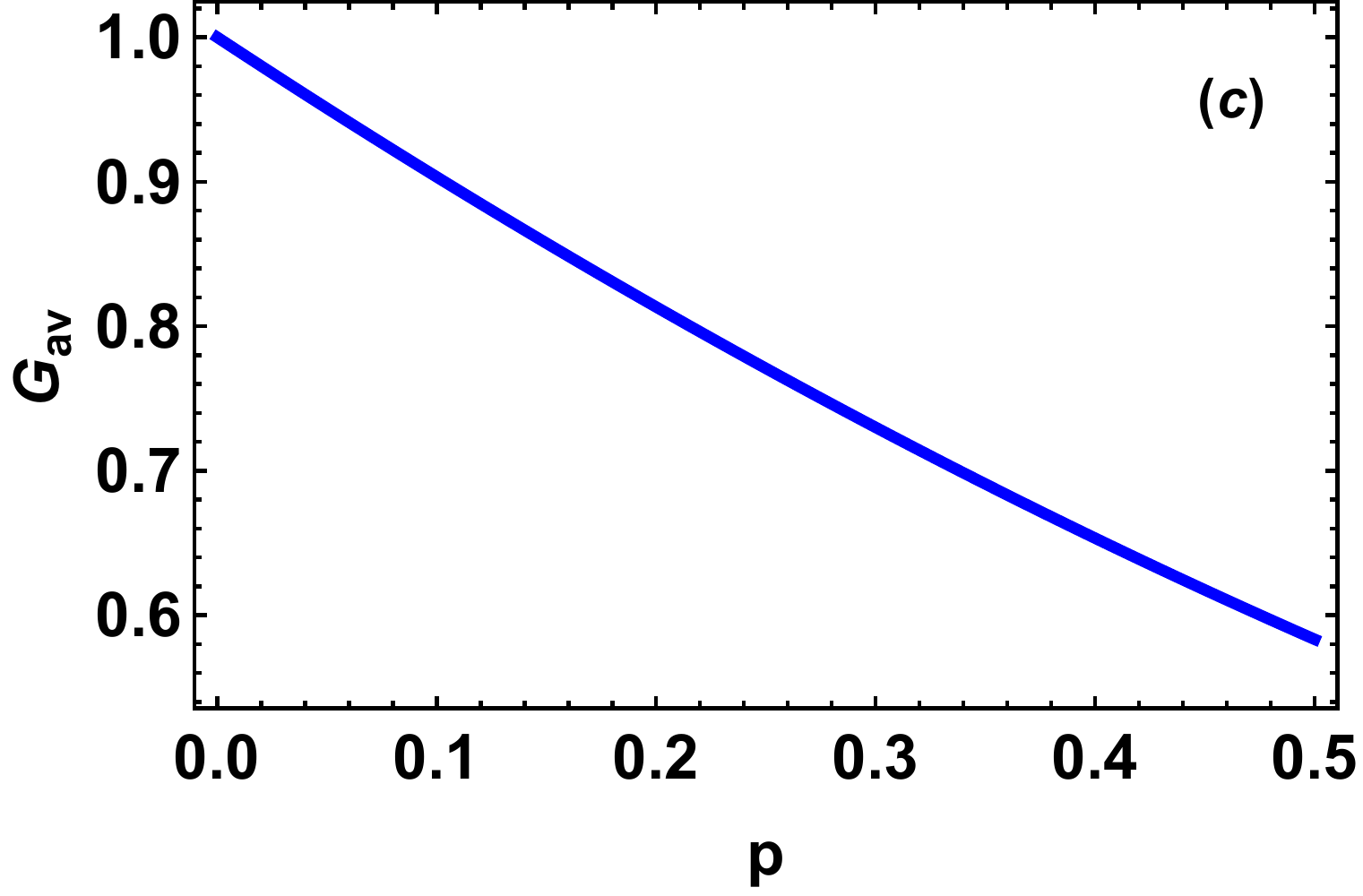}
		\includegraphics[width=43mm]{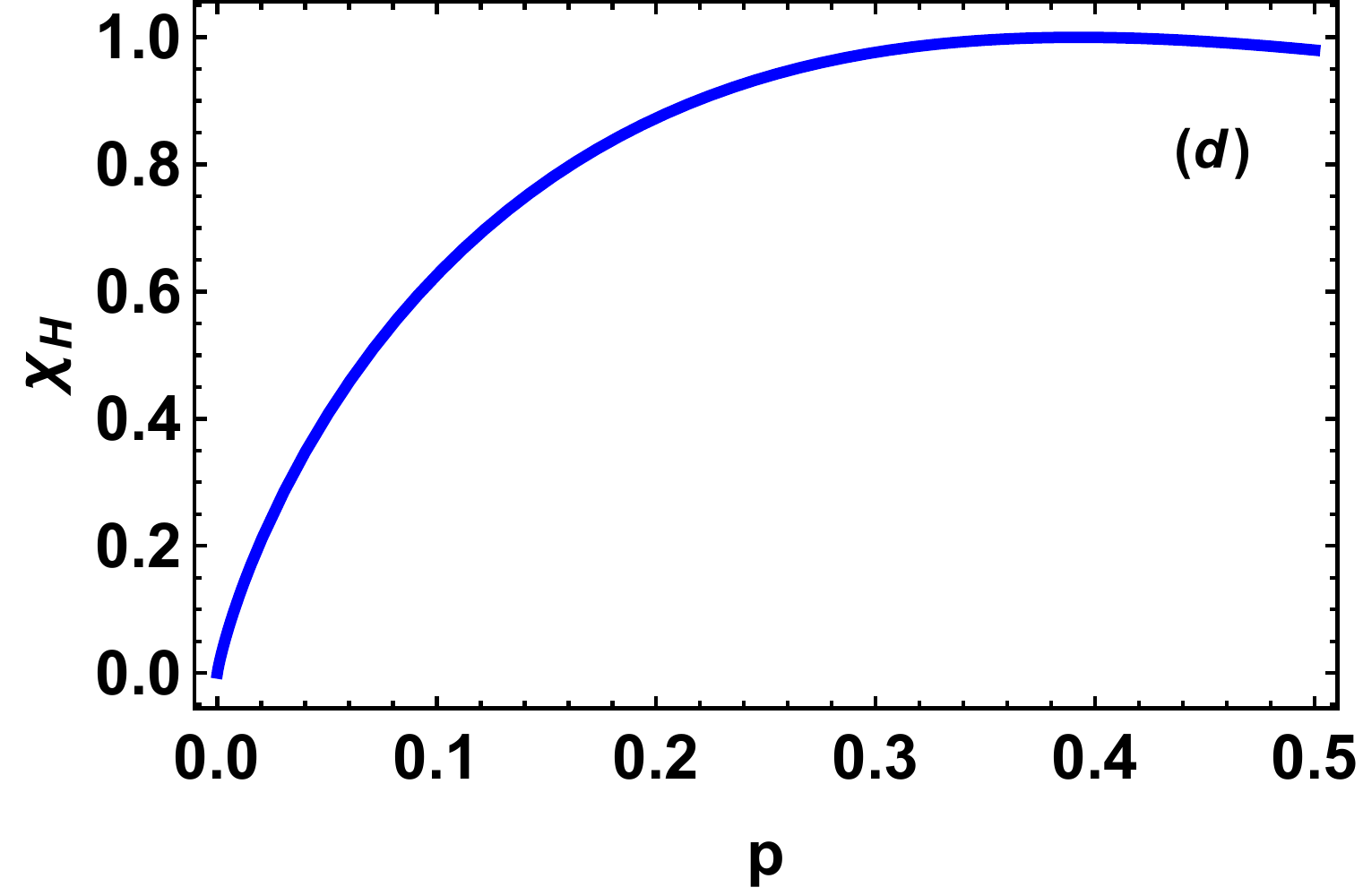}
	\end{center}
	\caption{Non-Markovina Dephasing (NMD) channel. Showing (a) coherence parameter (b) mixedness parameter (c) average gate fidelity and (d) Holevo quantity. The state parameters used are $\theta = \pi/4$ and $\phi =0$, and the channel parameter $\alpha =0.5$.} 
	\label{fig:joint_NMD}
\end{figure*}

\section{Results and discussion}\label{ResultsDiscussions}
The nature of the dynamics is governed by the decoherence rate, which is positive (negative) for Markovian (non-Markovian) dynamics. In the specific models considered in this work, namely RTN and NMD, the behavior of the respective decoherence rates is depicted in Fig. (\ref{fig:DecoRate}). This behavior is in concord with that seen with the QFI-flow. The non-Markovian behavior in case of RTN is controlled by the channel parameters, while the  NMD is non-Markoviann for all values of the parameter $\alpha$.  Figure (\ref{fig:RTN_Fisher}) depicts the QFI-flow corresponding to the state parameters $\theta$ and $\phi$ as a function of time.  The positive QFI-flow is a signature of non-Markovianity and is linked with the divisibility of the underlying dynamical map. It is well known that the non-Markovianity emerges in the RTN governed dynamics under the condition $2 a > \gamma$. In this regime, QFI-flow is found to oscillate symmetrically about zero, thereby confirming the non-Markovian nature of the dynamics. The behavior of the coherence and mixedness under RTN evolution is shown in Fig. (\ref{fig:Coh_Mix_RTN}). The coherence parameter $C$ and the mixedness parameter $\mathcal{M}$ decrease (increase) monotonically in the Markovian regime unlike the non-Markovian case. In the non-Markovian regime, these parameters shown recurrent behavior with time with an envelope of damped oscillation.  The interplay between  coherence and mixedness is symmetric in RTN model, Fig. (\ref{fig:Coh_Mix_Interplay}), such that the increase in one is accompanied with the decrease in other. The parameter $\beta$, Eq. (\ref{eq:beta}), depends only on the state parameter $\theta$, as given in Table (\ref{table}). Similar observations are made for the average gate fidelity ($G_{av}$) and Holevo quantity and are  depicted in Fig. (\ref{fig:Gfun_Holevo_RTN})   where the monotonic decrease with respect to time in Markovian case is contrasted with the oscillating behavior of these quantities in the non-Markovian scenario.\par
The decoherence rate in non-Markovian dephasing (NMD) model shows a negative branch separated from the positive branch by a singularity. However, the recurrent behavior observed in RTN is missing.  The  QIF-flow  is  positive  for certain range of the time like parameter $p$, as depicted in Fig. (\ref{fig:NMD_Fisher}), demonstrating the non-Markovian nature of this model.  The complementary behavior of coherence and mixedness is observed, that is, the decrease in the coherence is accompanied by an increase in the mixedness, Fig. (\ref{fig:Coh_Mix_Interplay}), such that the  $\beta$ parameter, defined in Eq. (\ref{eq:beta}), is a function of the state variable $\theta$, see Table (\ref{table}). The average gate fidelity decreases, while the Holevo quantity shows  an increase with $p$ in the range $0 \le p \le 1/2$.\\

\section{Conclusion}\label{Conclusion}
In this work, we considered two   quantum channels namely Random Telegraph Noise (RTN) and non-Markovian dephasing (NMD) and studied the dynamics of a general qubit state in these models. The dynamics is governed by completely positive and trace preserving  Kraus operators. The quantum Fisher information flow, which has recently been proposed as a witness of the non-Markovian behavior, is analyzed and is found consistent  with the analysis made using the decoherence rates in these models. Further,  various facets of quantum information viz., quantum coherence, mixedness, average gate fidelity and channel fidelity  are studied, their compact analytical expressions are obtained and their behavior is contrasted in the Markovian and non-Markovian regimes for the  RTN channel. Even though both RTN and NMD show non-Markovian behavior, there is a distinction between the two. The  non-Markovian dynamics for the RTN model has a characteristic recurrent behavior, not found in the case of NMD. Nevertheless, a symmetric trade-off between coherence and mixedness, quantified by a coherence-mixedness balance parameter $\beta$, is observed in both the cases, thereby testifying to their basic dephasing nature. Such characterization of the quantum channels can be significant from the perspective of  carrying out quantum information and communication tasks. 
\FloatBarrier
%
\end{document}